# Colossal terahertz emission with ultrafast tunability based on van der Waals ferroelectric NbOI$_2$

*Sujan Subedi[1], Wenhao Liu[2], Wuzhang Fang[1], Carter Fox[3], Zixin Zhai[2], Fan Fei[1], Yuan Ping[1,3,4], Bing Lv[2], Jun Xiao[1,3,5] \**

[1] Department of Materials Science and Engineering, University of Wisconsin-Madison, Madison, Wisconsin 53706, USA

[2] Department of Physics, University of Texas at Dallas, Richardson, Texas 75080, USA

[3] Department of Physics, University of Wisconsin-Madison, Madison, Wisconsin 53706, USA

[4] Department of Chemistry, University of Wisconsin-Madison, Madison, Wisconsin 53706, USA

[5] Department of Electrical and Computer Engineering, University of Wisconsin-Madison, Madison, Wisconsin 53706, USA

\*Corresponding author(s). E-mail(s): jun.xiao@wisc.edu

# Abstract

Terahertz (THz) technology is critical for quantum material physics, biomedical imaging, ultrafast electronics, and next-generation wireless communications. However, standing in the way of widespread applications is the scarcity of efficient ultrafast THz sources with on-demand fast modulation and easy on-chip integration capability. Here we report the discovery of colossal THz emission from a van der Waals (vdW) ferroelectric semiconductor $NbOI_2$. Using THz emission spectroscopy, we observe a THz generation efficiency an order of magnitude higher than that of ZnTe, a standard nonlinear crystal for ultrafast THz generation. We further uncover the underlying generation mechanisms associated with its large ferroelectric polarization by studying the THz emission dependence on excitation wavelength, incident polarization and fluence. Moreover, we demonstrate the ultrafast coherent amplification and annihilation of the THz emission and associated coherent phonon oscillations by employing a double-pump scheme. These findings combined with first-principles calculations, inform new understanding of the THz light-matter interaction in emergent vdW ferroelectrics and pave the way to develop high-performance THz devices on them for quantum materials sensing and ultrafast electronics.

# Introduction

The THz band of the electromagnetic spectrum is resonant with many low-energy collective excitations such as phonons, magnons, and superconducting Higgs modes [1-4]. This uniqueness opens new ways to selectively probe and control quantum materials with on-demand quantum properties, including dynamical multiferroicity [5], topological Floquet bands [6], and light-induced high-temperature superconductivity [7]. Moreover, ultrafast THz pulses with broad bandwidth could enable new ultrafast electronics and high-speed wireless communication with orders of magnitude improvement on the information capacity of the current technology [8]. To realize these exciting applications, it is crucial to generate ultrafast and intense THz waves with high efficiency and enable coherent control with high precision. One major approach to create ultrafast THz waves is based on optical rectification in organic and inorganic electro-optic crystals. For instance, organic crystals such as N-benzyl-2-methyl-4-nitroaniline (BNA) and *trans*-4′-

(dimethylamino)-*N*-methyl-4-stilbazolium tosylate (DAST), are renowned for their high electro-optic coefficients [9,10], but their instability in air and limited damage threshold greatly limits their practical applications [11-14]. While inorganic bulk crystals like ZnTe and LiNbO$_3$ provide greater robustness, they face severe challenges due to limited substrate compatibility, stemming from their 3D bulk bonding nature and significant lattice mismatch with many standard substrates such as GaAs, GaN and Si in the semiconductor industry [15-17]. This has hindered the development of THz integrated electronics and photonics, which are crucial for the long-sought advancements in ultrafast electronics and communications [18,19].

The emergent two-dimensional (2D) layered ferroelectrics may overcome the aforementioned challenges building upon their unique structural and electronic properties. First, 2D ferroelectrics are generally robust at room temperature. Furthermore, their van der Waals nature eliminates issues related to substrate lattice mismatch, enabling seamless integration with various substrates [20,21]. Many of these materials exhibit significant electrical polarization combined with a high nonlinear optical susceptibility [22]. In particular, recent advances show the great potential of niobium oxide dihalide NbOX$_2$ (X= Cl, I) semiconductors. NbOX$_2$ has a ferroelectric $C_2$ phase with monoclinic symmetry, where Nb cations exhibit non-zero off-center displacement along the Nb–O–Nb atomic chain direction, leading to one of the largest electrical polarizations (about 20 $\mu C/cm^2$) in attainable 2D ferroelectric [23]. Thanks to such a large ferroelectric polarization, recent calculations suggest a remarkably large linear electro-optic coefficient of about 300 pm/V (60 pm/V) under unclamped (clamped) conditions in ferroelectric NbOI$_2$ [24], which is 70(15) times larger than that in ZnTe [25] and 10(2) times larger than that in LiNbO$_3$ [26], suggesting its great potential for efficient ultrafast THz generation.

Here we demonstrate intense ultrafast THz emission with ultrafast tunability from the air-stable vdW layered ferroelectric NbOI$_2$ by both below- and above-bandgap ultrafast optical excitation. The corresponding THz generation efficiency is an order of magnitude higher than that in ZnTe, and the THz electric field is two orders higher than that in the nonpolar layered material 2H-MoS$_2$. The electric field of the generated THz wave through NbOI$_2$ can be up to about 0.1 MV/cm, which is comparable to that produced by optical rectification crystals like GaP and large electrical biased GaAs photoconductive emitter [27,28]. We further investigate the generation mechanisms by examining the dependence of THz emission on the NbOI$_2$ azimuthal angle, the polarization angle of the incident optical pump, and the optical pump fluence. In particular, we find the THz emission from below-bandgap excitation is predominantly governed by a non-resonant optical rectification mechanism associated with time-varying light-induced nonlinear polarization. In contrast, for the above-band gap excitation, additional contributions arise from transient orbital shift current in resonant optical rectification process and subsequent photocarrier drift current due to built-in ferroelectric polarization field. Building upon the above understandings, we employ a coherent double-pump scheme to demonstrate precise control over the THz emission and associated optical phonon oscillations in $NbOI_2$ at sub-picosecond scale to fully manipulate their phase and amplitude. Taken together, our findings highlight $NbOI_2$ as a promising candidate for colossal ultrafast THz generation and modulation, which is critical to advance THz band applications in quantum materials physics and high-speed electronics.

## Results and Discussion

Layered transition metal oxide dihalides are characterized by electronic and optical properties such as strong light-matter interaction [23], high charge carrier mobility [29], and tunable bandgaps [30], resulting from their highly anisotropic band structures and rich in-plane atomic arrangements. Among these materials, $NbOI_2$ exhibits a low-symmetry monoclinic structure formed by sheets of $NbO_2I_4$ octahedra. These octahedra are linked by the sharing of opposite iodine corners along the crystallographic c-axis and opposite oxygen corners along the b-axis, as illustrated in Figure 1(a). A second-order Peierl's distortion occurring along the b-axis involves the displacement of niobium ions from their central positions within the octahedral structure. The resulting asymmetric of the Nb-O bond lengths gives rise to spontaneous polarization along the b-axis. The spontaneous polarization in $NbOI_2$ is notably large (20 $\mu C/cm$) [23,31] and is essential to its giant nonlinear optical response. For instance, the effective nonlinear optical susceptibility $\chi^{(2)}$ such as $\chi_{yyy}$ ( $2d_{22}$ ) element along the polar axis has been reported to be exceptionally large ( ~ 160 pm/V at 1.7 eV) [32], with an order of magnitude greater than that of conventional nonlinear ferroelectric materials such as $BiFeO_3$, $BaTiO_3$, and $LiNbO_3$ [33-35]. We prepared $NbOI_2$ flakes with 1 mm lateral size from synthetic bulk crystals and conducted X-ray diffraction as well as Raman measurements to confirm the desired crystalline (see Methods and Supplementary Section S1). Indeed, our polarization-dependent second-harmonic generation (SHG) (Figure 1(b)) reflects the anisotropy with large enhancement along the polar axis, in contrast to the isotropic SHG response observed in the higher symmetry semiconductor 2H-MoS₂ (space group $D_{6h}$) [36]. Such large optical nonlinearity also enables ultrathin NbOI₂ to efficiently generate quantum entanglement photon sources via a spontaneous parametric down-conversion process [37,38].

Beyond the visible spectrum, the combined large ferroelectric polarization and strong light-matter interactions in NbOI₂ are expected to allow substantial ultrafast THz emission originating from nonlinear optical rectification and the transient injection currents. Moreover, NbOI₂ has an indirect band gap $E_g$ around 1.7 eV [29]. This semiconductor nature is favorable for efficient THz sources, which bypasses the unwanted optical and THz absorption issues in metallic THz emitters [39]. Inspired by these attributes, we investigate the THz wave generation from NbOI₂ by ultrafast THz emission spectroscopy in a transmission configuration. The electric field $E_{THz}$ of THz emission is detected using the standard electro-optical sampling (EOS) technique. This technique can provide both amplitude and phase information of the emitted signal. The THz emission schematic is outlined in Figure 1(c), and detailed experimental setup information can be found in Supplementary Section S2. We compare the emitted THz waveform and corresponding fast Fourier transform (FFT) for approximately 300 $\mu$m thick NbOI₂ and 0.5 mm thick ZnTe under 400 nm excitation with the same experimental conditions (Figure 1(d)). We find the peak-to-peak THz electric field intensity from NbOI₂ to be about 1.5 times of that from ZnTe, one of the major THz generation nonlinear crystals. A maximum peak electric field is estimated to be close to 0.1 MV/cm (see Supplementary Section S3). To quantify how efficient the THz generation is, we calibrate and compare the THz emission efficiency $\eta_{THz}$ among NbOI₂, ZnTe, and nonpolar vdW 2H-MoS₂ semiconductors. It is an important metric that quantifies the intrinsic ability of a material without certain sample geometric influence to convert optical excitation into THz radiation. We evaluate the THz amplitude per unit sample thickness (THz emission efficiency) by the following expression [40],

$$\eta_{THz} = \frac{E_{THz}}{nF\delta} \quad \dots\dots\dots\dots\dots\dots\dots\dots\dots\dots\dots\dots\dots\dots\dots\dots\dots\dots\dots(1)$$

where $E_{THz}$ is the THz electric field for electro-optical sampling, $n$ is the refractive index in the THz field, $F$ is the fluence, and $\delta$ is the penetration depth. For NbOI₂ and ZnTe, we measure the reflection coefficient at 400 nm at normal incidence and apply the Fresnel equations to calculate the penetration depth ($\delta$). For 2H-MoS$_2$, the penetration depth is calculated based on the reported absorption coefficient (see Supplementary Information Section S3 for more information). The calculated THz emission efficiency for the three materials at a pump fluence of 5.6 mJ/cm² is shown in Figure 1(e). We find the THz emission efficiency of NbOI$_2$ is approximately an order of magnitude greater than that of 0.5 mm-thick ZnTe, and about 20 times stronger than 2H-MoS$_2$. This enhanced performance can be attributed to the significantly higher electro-optical and nonlinear optical coefficients in NbOI$_2$ compared to ZnTe and 2H-MoS$_2$ [35,37,38]. Note that the measured cutoff frequency of THz emission from NbOI$_2$ can be up to 4.0 THz using a thinner <110> ZnTe crystal (0.5-mm thick) for EO sampling (Supplementary Section S4). Further, The THz emission from the NbOI$_2$ is found to be air-stable with negligible emission feature change over one-month exposure upon ambient condition (Supplementary Section S5).

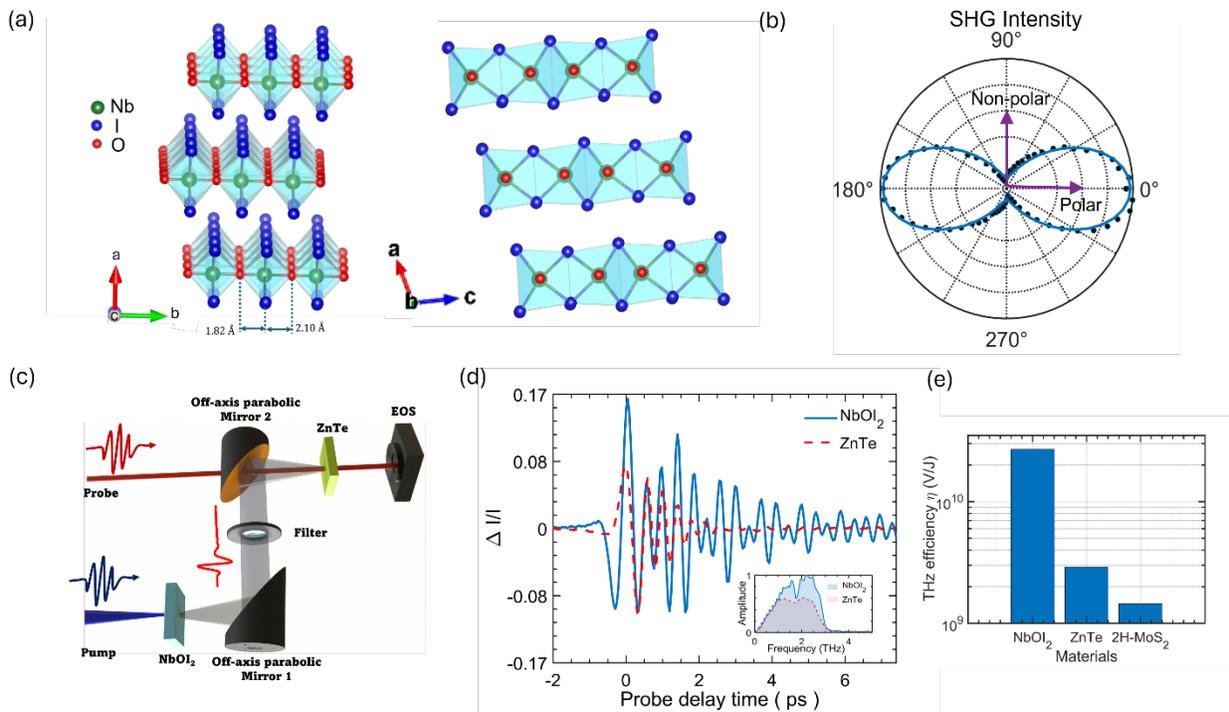

**Figure 1: Colossal THz emission from layered ferroelectric semiconductor NbOI$_2$.** (a) A three-dimensional schematic diagram of the top and side view of the NbOI$_2$ crystal structure. In the schematic, niobium (Nb) atoms are in green, iodine (I) atoms are in blue, and oxygen (O) atoms are in red. The crystal structure of NbOI$_2$ is based on the fundamental building block of NbO$_2$I$_4$ octahedra. The in-plane ferroelectricity arises from the polar displacement of Nb atoms relative to the center of the NbO$_2$I$_4$ octahedra

along the crystallographic b axis. The in-plane polarization is parallel aligned for each layer of NbOI$_2$, where all layers are interconnected through the weak interlayer vdW force. (b) Polar plots showing the total second harmonic generation (SHG) intensity as a function of the polarization angle of the excitation beam at $\lambda_{exc}$ = 800 nm. The black dots represent the experimental data, which match well with the corresponding fitting based on the crystal space group (the blue curve). (c) The schematic diagram for the THz time domain emission spectroscopy setup (THz-TDES) used to measure the THz emission of the thick NbOI$_2$ samples. (d) Waveforms of the THz radiation generated using a laser wavelength of $\lambda_{exc}$ = 400 nm for NbOI$_2$ and ZnTe under same experimental conditions and the inset shows their FFT amplitude. The peak-to-peak THz emission intensity from NbOI$_2$ is 1.5 times stronger than that from ZnTe, with a maximum peak electric field close to 0.1 *MV/cm* (see Supplementary Section S3). (e) Comparison of THz emission efficiency between different typical materials. The THz emission efficiency for NbOI$_2$ is about 10 times stronger than ZnTe and about 20 times stronger than the nonpolar semiconductor 2H-MoS$_2$.

To understand the generation mechanisms of the observed colossal THz emission from NbOI$_2$, we first investigate the polarization state of the emitted THz wave and its dependence on the optical pump polarization. To implement, we fix the polar axis of the sample parallel to the plane of the lab frame and use a half-wave plate to rotate the linear polarization angle of the 400 nm pump (Figure 2a). We separately probe the $E^x_{THz}(E^y_{THz})$ component of the THz field by keeping the EOS probe polarization along the horizontal axis, the ZnTe EO sampling crystal (001) direction along vertical (horizontal) axis and the THz wire grid polarizer along horizontal (vertical) axis. The peak-to-peak intensity of the individual $E^x_{THz}$ and $E^y_{THz}$ components as a function of incident polarization angle of the above bandgap (400 nm) excitation are shown in Figure 2(b). Both of them show much larger emission intensity when optical pump polarization is along the polar axis. This excitation anisotropy indicates that spontaneous polarization along the polar axis enhances the asymmetry in the material's response to excitation. Besides, the $E^x_{THz}$ component (parallel to the polar axis) is about six times larger than the $E^y_{THz}$. Such an emission anisotropy again suggests a strong dependence of THz emission on the material's intrinsic polarization along the polar axis. Another finding is that the THz emission is linearly polarized without phase retardance between the $E^x_{THz}$ and $E^y_{THz}$ components revealing by their time trajectories (Figure 2c). We also explore how the sample's azimuthal angle-dependent THz emission pattern to elucidate the influence of crystallographic symmetry on emission characteristics. To study the azimuthal angle dependence, we fix the incident polarization horizontal to the lab frame and rotate the sample. The polarization of the THz emission is analyzed by measuring both the $E_{THz}^x$ and $E_{THz}^y$ components. Figure 2(d) shows the polarization angle of the THz emission ($\psi$) as a function of the azimuthal angle ($\phi$). The observed $\psi=\phi$ relation confirms the ferroelectric polarization play a critical role in THz emission mechanisms. Thus, the THz waveform can be flipped when the ferroelectric spontaneous polarization is reversed achieved by 180-degree azimuthal sample rotation (Figure 2e). The full peak-to-peak THz emission dependence on the azimuthal angle is shown in Supplementary Section S6.

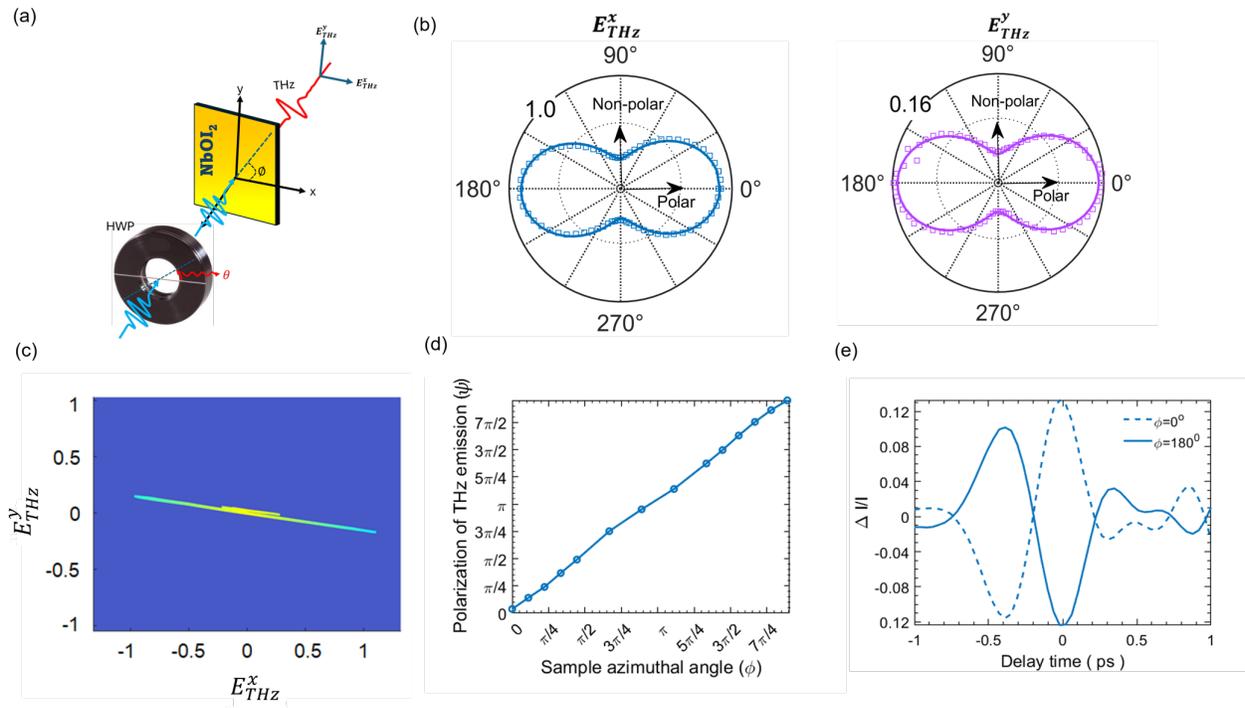

**Figure 2: Highly anisotropic THz emission pattern of layered ferroelectric NbOI₂.** (a) Schematics for the polarization and azimuthal angle dependent THz emission. (b) Peak-to-peak $E_{THz}^x$ and $E_{THz}^y$ components of the THz electric field as a function of the polarization angle of pump excitation at 400 nm when the polar axis is parallel to the horizontal component of the lab frame. The $E_{THz}^x$ THz emission is approximately 6 times stronger than the $E_{THz}^y$ component. Data are fitted for the THz emission from $C_2$ space group materials. (c) Corresponding time trajectory of emitted THz field, indicating linear polarized emission nature. (d) Dependence of the angle of polarization of terahertz emission ($\psi$) on the azimuthal rotation angle ($\phi$) of the sample with normal pump incidence. The observed relation $\psi=\phi$ is consistent with the space group $C_2$ symmetry. (e) Azimuthal angle dependence of the waveform of the THz electric field, where the electric-field transients reverse their sign at crystal azimuthal angles of 0° (dashed line) and 180° (solid line).

Building upon the significance of ferroelectric polarization in observed THz emission, we further study the fluence dependent THz emission upon below and above bandgap excitation to reveal the THz emission mechanisms in NbOI$_2$ (Figure 3(a), (b), (c)). These two scenarios can distinct the time-varying ferroelectric polarization and transient current contributions, as detailed below. Firstly, we observe that the peak-to-peak value for the above bandgap excitation and FFT spectrum is 1.5 times greater than below bandgap excitation (Figure 3(a) and inset), which indicates that THz emission above band gap involves additional contributions which will discuss below. Secondly, the fluence-dependent THz emission trend is also distinct with linear and saturation behaviors, respectively (Figure 3(b)). Thirdly, THz emission patterns from both cases exhibit a two-fold symmetry (Figure 3(c)). These results highlight the differences and similarities in the THz generation mechanisms linked to the two types of optical excitations. For below bandgap excitation (800 nm), there are no photocarriers are excited and the only THz emission source is the time-varying ferroelectric polarization $P$ (first panel of Figure 3d), induced by the non-resonant optical rectification process. This direct ferroelectric polarization origin explains the observed anisotropic emission pattern (Figure 3(c)). Namely, $E_{THz}^x(t) = \frac{\partial^2 P(t)}{\partial^2 t} = \chi^{(2)} E(\omega_o) E(-\omega_o + \omega_{THz}) = \chi^{(2)} I(t)$, where $\chi^{(2)}$ represents the second-order susceptibility tensor for optical rectification and $I(t)$ denotes the optical excitation intensity. Since $I(t)$ is proportional to the incident fluence, the THz electric field intensity thus shows a linear dependence on the excitation fluence for 800 nm pump across a wide range of pump fluence, extending up to 10.3 mJ cm$^{-2}$. In contrast, for 400 nm excitation, additional THz emission contributions take place leading to observed larger emission strength, including transient photocurrent $J_{shift}$ induced by the resonant optical rectification and transient drift current $J_{drift}$ arises from the acceleration of photo-excited carriers under the polarization-induced in-plane electric field. In particular, the conduction band (CB) is predominantly comprised of Nb 4$d$ states, whereas the maxima of the valence band (VB) are primarily formed from mixing of I 5$p$ and Nb 4$d$ states [41]. As a result, upon above bandgap excitation, the excited photocurrent shifts towards the Nb atoms. This shift of the spatial charge center from the VB to the CB produces a transient nonlinear photocurrent $J_{shift}$ to emit additional THz radiation (second panel of Figure 3d). Meanwhile, extra THz emission can come from the drift current $J_{drift}$ arises from the separation and acceleration of photoexcited carriers driven by spontaneous polarization field along the polar axis. (third panel of Figure 3d).

On the other hand, at high fluence excitation ferroelectric spontaneous polarization generated surface photocurrent can lead to partial ferroelectric depolarization [42]. Specifically, when m bound carriers per unit area are excited, the spontaneous polarization decreases and is represented by $P_f = (n - m)p$, which adverse the further increment of THz emission. To quantify the contributions from different mechanisms, we model the saturation trend upon 400 nm excitation by the following equation [43].

$$E_{THz}^x = S_1 F + S_2 \left[1 - \frac{1}{\left(1+\frac{F}{F_s}\right)}\right] \dots \dots \dots \dots \dots \dots \dots \dots \dots \dots \dots \dots \dots \dots \dots (2)$$

Here, $S_1$ is the contribution of the $J_{shift}$ and $J_{drift}$ obtained from resonance optical rectification along with spontaneous polarization field which is linear with the incidence intensity, $S_2$ defines the saturation behavior due to partial ferroelectric depolarization due to ferroelectric spontaneous polarization generated photocurrent which screen the THz radiation, and $F_s$ represents the saturation fluence. The above bandgap excitation fits well with the model function Equation [2], as shown in Figure 3b indicating a saturation fluence 6.8±1.8 mJ cm$^{-2}$ upon 400 nm excitation.

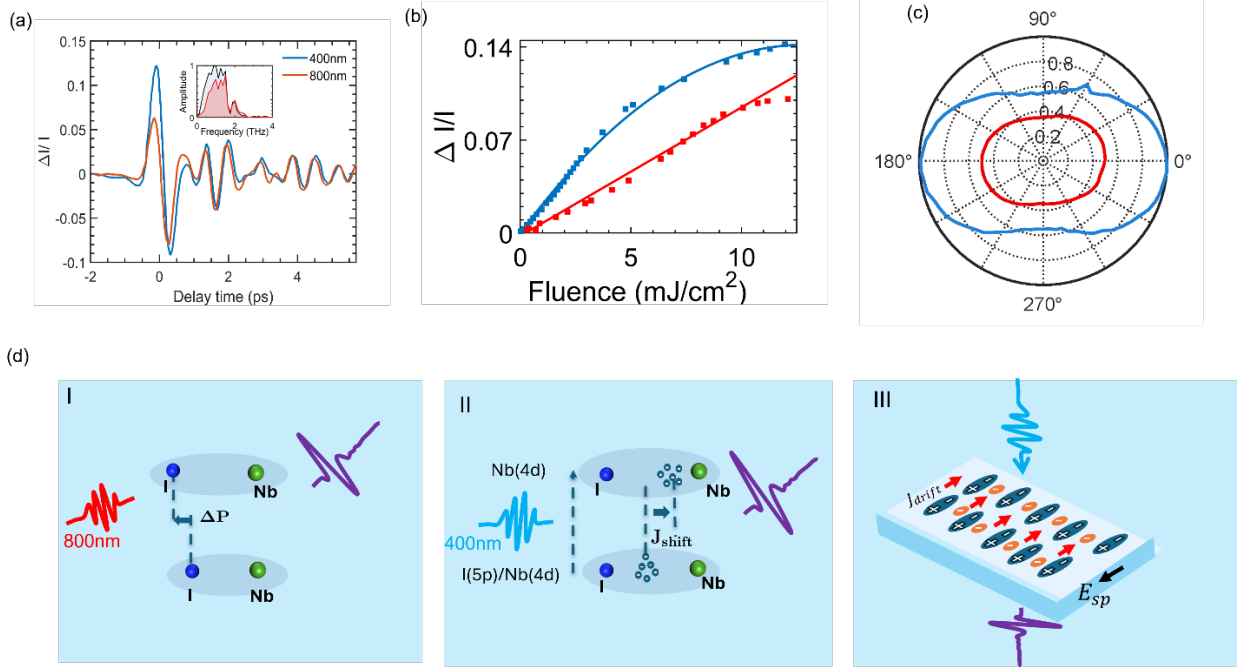

**Figure 3: THz emission mechanisms in layered ferroelectric NbOI$_2$.** (a) Comparison of the magnitudes of the time-domain THz traces from below band gap excitation $\lambda_{exc}$ = 800 nm (red) and above band gap excitation $\lambda_{exc}$ = 400 nm (cyan). (b) Dependence of the THz signal magnitude generated by NbOI$_2$ on pump fluence for below and above band gap excitation. The solid line is a fit to Equation [2]. (c) Polar plot of the peak-to-peak $E_{THz}^x$ component of the THz field as a function of the polarization angle of the pump excitation at 800 nm (red) and 400 nm (cyan), with the polar axis aligned parallel to the horizontal component of the lab frame. (d) Schematic representation of THz pulse emission mechanisms: first panel illustrates induced polarization ($\Delta P$) via non-resonant optical rectification in NbOI$_2$ under below band gap excitation ($E_{exc} < Eg$), while for the above bandgap excitation ($E_{exc} > Eg$) the second panel illustrates the shift current due to resonant optical rectification effect and the third panel shows the drift current driven by spontaneous polarization.

Besides the discovery of colossal THz emission, we also observed the presence of long-lived coherent phonon oscillations following the initial transient signal (Figure 4(a)). To analyze the dynamics of this oscillatory behavior, the FFT of the time-domain THz signal was performed after removing the initial transient component (the inset of Figure 4(a)). It reveals two specific spectral peaks at 1.78 THz and 2.86 THz, with a bandwidth of about 0.2 THz and 0.35 THz, respectively. Additionally, the long scan of the time-domain THz signal shows the lifetime of the oscillations extends up to 20 ps (see Supplementary Section S7 for long scan data and Gaussian fit of the spectral peaks). To understand the nature of those long-lived coherent phonons, we conducted first principle calculations and the detailed results are presented in section S8 of the Supplementary Information. The 1.78 THz peak is attributed to the motion of Nb-O atoms along the polarization axis, while I atoms move in the opposite direction (Figure 4(b), Supplementary Section S8 and Supplementary video 1). As for the $f_2 = 2.86$ THz mode, it is attributed to a relative motion between the Nb and O atoms, resulting in dynamic changes of spontaneous polarization in $NbOI_2$ (Figure 4(b), Supplementary section S8 and Supplementary video 2). These assignments are consistent with the fact that phonon-mediated THz emission oscillation primarily results in the $E_{THz}^x$ component (See Supplementary Section S9) suggesting the involved ionic motions along the same axis (Nb-O polar axis).

Such long-lived phonons allow well-defined coherence in a few oscillation cycles and can be utilized for on-demand fast modulation, another long-sought attribute in ultrafast THz sources. Specifically, we use a double-pump scheme to achieve coherent control of the THz emission and subsequent coherent phonon oscillations in $NbOI_2$ crystal (Figure 4(c)). Two excitation pulses with the same fluence and polarization are employed. The interval between the two pulses ($\tau$) is fine-tuned using a motorized delay stage. Thanks to the long-lived phonon coherence, we are able to superimpose the THz emission oscillations induced by both the first and second pulses. The phase difference between these oscillations is determined by $2\pi f_1 \tau$ and $2\pi f_2 \tau$. When $2f_1\tau$ ($2f_2\tau$) is an even integer, constructive interference occurs and amplify the $f_1$ ($f_2$) phonon oscillation. Conversely, when $2f_1\tau$ ($2f_2\tau$) is an odd integer, destructive interference takes place for the $f_1$ ($f_2$) mode, resulting in the cancellation of the THz emission oscillation. In Figure 4(d), we present a 2D colormap plot showing the THz emission dependence of the double-pump delay time $\tau$ and the probe delay time. The amplitude and phase of the THz signal following excitation by the second pulse exhibits periodic modulation as a function of the double-pump delay time $\tau$. For instance, the blue curve of Figure 4(e) shows the amplified THz emission for $\tau = 2.8$ ps, which meets the constructive interference condition for both the 1.78 and 2.86 THz phonon modes. Conversely, the red curve illustrates the annihilated THz emission for $\tau = 2.6$ ps, which satisfies the destructive interference condition for these two phonon modes. For the FFT spectrum of constructive and destructive interference conditions at these delay time periods, see Supplementary Section S10. Figure 4(f) summarizes the amplitude and phase of the 1.78 THz (top) and 2.86 THz (bottom) oscillation modes with the varying delay time between the pump pulses following constructive interference when $2f\tau$ is an even integer and destructive interference when $2f\tau$ is an odd integer. In the top panel, the red dots are the FFT amplitude for the different pump delays for coherent oscillation of 1.78 THz. The red solid line is a fit to an oscillatory function. Similarly, the bottom panel shows the data (cyan dot with dash line) and fit (solid line) for the 2.86 THz mode. These results demonstrate the full control capability of ultrafast THz emission amplitude and phase, which is critical for on-demand THz signal processing in ultrafast electronics and coherent control of THz collective excitations in quantum materials.

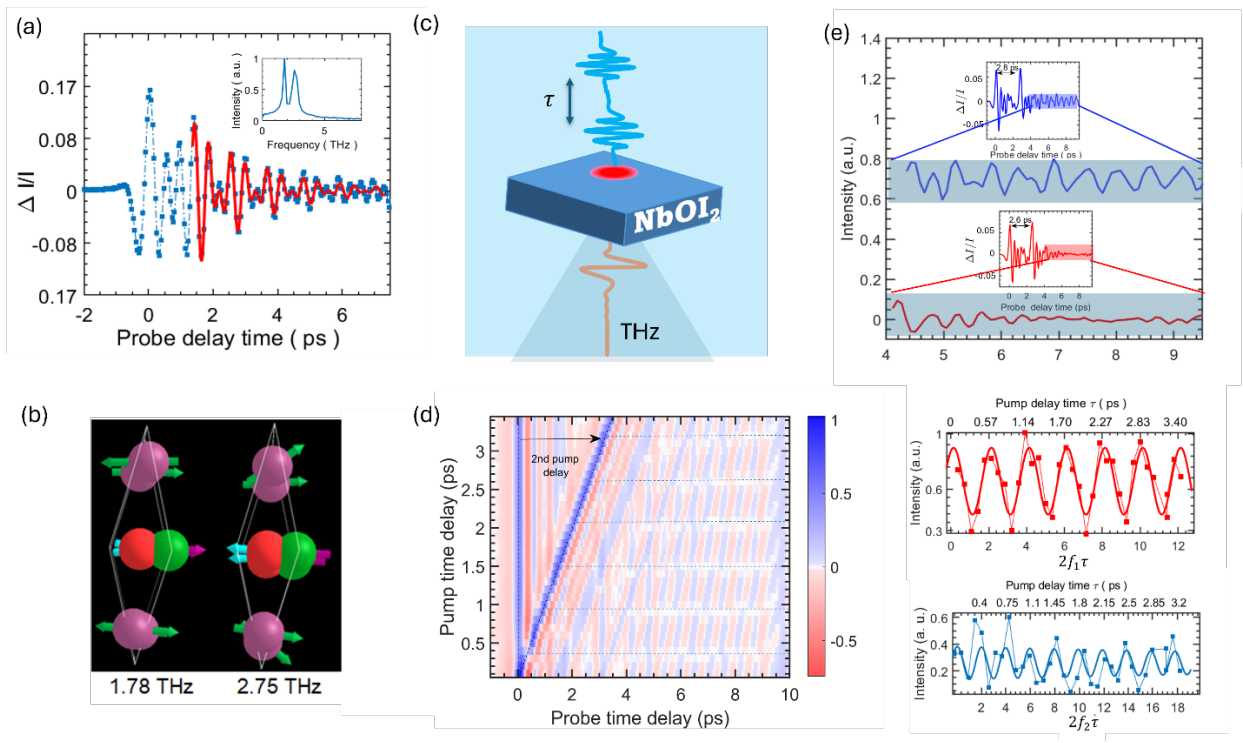

**Figure 4: Ultrafast coherent control of THz emission and phonon oscillations in NbOI$_2$.** (a) The THz emission signal from NbOI$_2$ displays long-lived coherent oscillations (dashed line). The red curve shows a fit of the long-lived oscillatory part of the emission to a decaying double cosine function, yielding two mode frequencies of 1.78 THz and 2.86 THz. The inset is the corresponding the FFT spectrum. (b) Density functional theory calculated vibrational modes of the 1.78 and 2.75 THz phonons. The arrows indicate the direction of vibration. (c) The schematics of ultrafast coherent control of THz emission by a double-pump scheme. (d) 2D color map of THz emission dependence of the double-pump delay time $\tau$ and the probe delay time. Periodic modulation of THz emission amplitude and phase are demonstrated. (e) Typical constructive (e.g., $\tau$ = 2.8 ps, top blue part) and destructive interference (e.g., $\tau$ = 2.6 ps, bottom red part)

of THz emission and associated coherent phonon, respectively. (f) The amplitude and phase of the 1.78 THz (top) and 2.86 THz (bottom) oscillation modes controlled by the varying double-pump delay time $\tau$.

## Conclusions

In summary, we report the discovery of colossal ultrafast THz emission from $NbOI_2$ upon either below or above bandgap excitation. The THz emission efficiency can be up to an order of magnitude greater than that of ZnTe. For below bandgap excitation (800 nm), the intense THz emission originates from the nontrivial non-resonant optical rectification mediated by the large ferroelectric polarization. For the above bandgap optical excitation (400 nm), the THz emission is further enhanced by resonant optical rectification combined with large drift currents induced by the in-plane ferroelectric polarization. Moreover, we have demonstrated ultrafast and precise modulation of the amplitude and phase of the THz emission and coherent phonon oscillations by utilizing coherent double-pump excitations. Distinct from prior THz emission studies in layered semimetals such as $PtSe_2$ [40] and $WTe_2$ [44], our results have unique significance in terms of: (1) the semiconductor nature of $NbOI_2$ bypasses the notorious optical and THz absorption issues in layered metals, which leads to much larger THz electric field, (2) the superior air stability of $NbOI_2$ under ambient conditions, in contrast to air-sensitive $WTe_2$ [45], and (3) the long-lived and narrow bandwidth phonons in $NbOI_2$ allow for the demonstrated coherent ultrafast modulation of THz emission with precise amplitude and phase control. Taken together, our study advances the understanding of strong THz light matter interactions in 2D vdW ferroelectrics and paves the way for using low-dimensional intense THz sources with ultrafast tunability for quantum material sensing, ultrafast electronics and next-generation wireless communications.

**Experimental Section**

*Crystal Growth and sample Preparation:*

NbOI$_2$ single crystals were synthesized using the chemical vapor transport method. Niobium powder (99.8%, Alfa Aesar), powder (99.9%, Alfa Aesar) and Iodine flakes (99.99%, Alfa Aesar) were mixed in an appropriate ratio under an inert atmosphere. The mixtures were loaded into a silica tube, which was flame-sealed under vacuum. To reduce the loss of Iodine during sealing, liquid nitrogen was utilized. The ampoule was then loaded in a horizontal tube furnace with the hot end containing the reactants maintained at 700°C and the cold end at 650°C. After 10 days of reaction, plate-like single crystals with size of several millimeters formed at the cold end. Large flat NbOI$_2$ samples of ~2x2 mm size and 300 $\mu$m thick are cleaved from bulk crystals and placed on Scotch Tape that has a ~1 mm diameter hole punched through, providing a ~1 mm size flat area that is significantly larger than the beam size.

*Second harmonic generation:*

A second harmonic generation (SHG) experiment was performed using femtosecond laser pulses from a Ti:Sapphire laser, operating at a central wavelength of 800 nm, with a pulse duration of 35 fs and a repetition rate of 1 kHz. The polarization of the incident 800 nm light was adjusted using a half-wave plate before being focused onto the sample surface, resulting in an approximate spot size of 500 μm and a fluence of 1.2 mJ/cm². A collecting lens, configured in a reflection geometry, captured the emitted SHG signal from sample, which was then filtered through a optical bandpass filter prior to detection with a photomultiplier tube (PMT). The SHG signal recorded by the PMT was analyzed using a lock-in amplifier to minimize noise interference.

*THz emission measurement:*

For the THz emission spectroscopy setup for NbOI$_2$, we used the femtosecond laser pulses emitted from a Ti:Sappire laser with central wavelength at 800 nm, a pulse duration of 35 fs, and a 1-kHz repetition frequency. The beam from the laser output is split into two arms using a beam splitter for pump and probe beams. The 400 nm beam generated from the BBO crystal and 800 nm light from the laser output are used as pump beams to excite the sample at normal incidence. The spot size of pump beam is fixed to 225 $\mu$m. The half-waveplate is used to change the polarization of the pump beam. The probe beam is used to detect the THz radiation by electro-optical sampling (EOS) method. For EOS, we used a <110> ZnTe crystal with a thickness of 1mm to measure two orthogonal components of the electric field of a terahertz pulse. We measure the horizontal (vertical) THz component, which is parallel (normal) to the experimental table and normal to the THz propagation direction. To measure the horizontal and vertical components, we rotate the ZnTe crystal and wire grid polarizer while keeping the EOS probe polarization horizontal. When the ZnTe (001) direction is vertical (horizontal), only the horizontal (vertical) component of THz is detected. As for the double pump-probe experiment, a single pump beam is split into two pump beams of equal power using a beam splitter. Both beams pass through a BBO crystal to generate 400 nm excitation beams. The emitted beams are collimated again using a beam splitter and then focused to a spot size of approximately 225 μm, with a fluence of 5.3 mJ/cm² at the sample for THz emission. The time delay between the two beams is controlled using a translational stage on one of the pump beams. The emitted THz field is measured using the EOS method, as in the single-pump THz emission experiment.

*First-principle calculations:*

Density functional theory calculations were performed using the Vienna Ab Initio Simulation Package (VASP) [46]. The projector augmented wave (PAW) method [47] implemented in VASP was used with a kinetic energy cutoff of 600 eV. The generalized gradient approximation of Perdew-Burke-Ernzerhof (PBE) [48] was used as the exchange-correlation functional. A primitive cell of 8 atoms (2Nb, 4I, 2O) was used. A Γ-centered 12x12x6 k-point mesh was used for integration in the first Brillouin zone. Structure was fully relaxed until the force on each atom is less than 0.005 eV/Å. DFT-D3 method was used to account for the dispersion correction [49]. Phonon properties were calculated using the Phonopy code [50]. A 3x3x3 supercell was used to calculate the interatomic force constants based on the finite difference method. Phonon properties were obtained by diagonalizing the dynamical matrix, which is built on interatomic force constants.

**Acknowledgements**

*S.S., W. F., F.F., Y. P., and J.X. gratefully acknowledge primary support by NSF through the University of Wisconsin Materials Research Science and Engineering Center (DMR-2309000). C. F. acknowledges additional support from the U.S. National Science Foundation (DMR-2237761). J.X. acknowledges additional support from the Office of Naval Research (N00014-24-1-2068). Work at UT Dallas is supported by the US AFOSR Grant No. FA9550-19-1-0037 & FA9550-21-1-0297, NSF-DMREF-2324033, and ONR Grant No. N00014-23-1-2020. This research used resources of the National Energy Research Scientific Computing Center (NERSC) a U.S. Department of Energy Office of Science User Facility operated under Contract No. DE-AC02-05CH11231, and the resources of Center for High Throughput Computing. (2006). Center for High Throughput Computing. doi:10.21231/GNT1-HW21.*

**Author Contributions**

J.X. and S.S. conceived the research and designed the experiments. J.X. supervised the project. W.L. and Z.Z. synthesized the bulk high-quality $NbOI_2$ crystals under the guidance of B.L.; C.F., and F.F. fabricated the large and flat $NbOI_2$ films under the guidance of J.X.; S.S. performed the THz emission and SHG measurements and analyzed the data with J.X.; W. F. and Y. P. performed first-principles calculations. All authors discussed the results and jointly wrote the paper.

**Data Availability**

The data that support the findings of this study are available from the corresponding author upon reasonable request.

**Competing Interests**

The authors declare no competing interests.